\documentclass[proof]{WileyASNA-v1}

\articletype{Article Type}%

\received{26 April 2016}
\revised{6 June 2016}
\accepted{6 June 2016}

\begin{document}

\title{Tidal disruption events and quasi periodic eruptions}

\author[1]{Natalie A. Webb}

\author[1]{Didier Barret}

\author[1]{Olivier Godet}
\author[1]{Maitrayee Gupta}

\author[2]{Dacheng Lin}
\author[1]{Erwan Quintin}

\author[1]{Hugo Tranin}

\authormark{Webb \textsc{et al}}

\address[1]{\orgdiv{Universit\'e de Toulouse, CNRS, CNES}, \orgname{Institut de Recherche en Astrophysique et Plan\'etologie}, \orgaddress{\state{Toulouse}, \country{France}}}
\address[2]{\orgdiv{Space Science Center}, \orgname{University of New Hampshire}, \orgaddress{\state {Durham}, \country{USA}}}

%\address[2]{\orgdiv{Org Division}, \orgname{Org Name}, \orgaddress{\state{State name}, \country{Country name}}}

%\address[3]{\orgdiv{Org Division}, \orgname{Org Name}, \orgaddress{\state{State name}, \country{Country name}}}

\corres{*\email{Natalie.Webb@irap.omp.eu}}

%\presentaddress{This is sample for present address text this is sample for present address text}

\abstract{Tidal disruption events (TDEs) occur when a star passes close to a massive black hole, so that the tidal forces of the black hole exceed the binding energy of a star and cause it to be ripped apart. Part of the matter will fall onto the black hole, causing a strong increase in the luminosity. Such events are often seen in the optical or the X-ray (or both) or even at other wavelengths such as in the radio, where the diversity of observed emission is still poorly understood. The XMM-Newton catalogue of approximately a million X-ray detections covering 1283$^2$ degrees of sky contains a number of these events. Here I will show the diverse nature of a number of TDEs discovered in the catalogue and discuss their relationship with quasi periodic eruptions.}

\keywords{X-rays: galaxies, galaxies: nuclei, galaxies: dwarf, catalogs}

%\jnlcitation{\cname{%
%\author{Williams K.}, 
%\author{B. Hoskins}, 
%\author{R. Lee}, 
%\author{G. Masato}, and 
%\author{T. Woollings}} (\cyear{2016}), 
%\ctitle{A regime analysis of Atlantic winter jet variability applied to evaluate HadGEM3-GC2}, \cjournal{Q.J.R. Meteorol. Soc.}, \cvol{2017;00:1--6}.}

%\fundingInfo{This work was funded in part by the CNES and the European Union's Horizon 2020 research and innovation programme under grant agreement n°101004168, the XMM2ATHENA project.}

\maketitle

%\footnotetext{\textbf{Abbreviations:} ANA, anti-nuclear antibodies; APC, antigen-presenting cells; IRF, interferon regulatory factor}

\section{Introduction}
\label{sec:intro}

Whilst many supermassive ($\sim$10$^{6-10}$ M$_\odot$) black holes (SMBH) are known in the cores of massive galaxies, it is still not clear how they form, nor how they evolve. It is unlikely that SMBH form from stellar mass black holes, as even accreting continuously at or above the Eddington limit (the maximum rate for material to be accreted onto the black hole supposing spherical accretion), it is difficult to reach masses as high as $\sim$10$^9$ M$_\odot$ as early as z$\sim$7.1 \citep{mort11} or even 8$\times$10$^8$ M$_\odot$ at z=7.54 \citep[0.69 Gyr,][]{bana18}. It has been proposed that they may form from lower mass black holes, known as seed black holes, or intermediate mass black holes (IMBH, 10$^{2-5}$ M$_\odot$), but few of these objects have been found. There may also be a mechanism to accrete above the Eddington limit, thus allowing SMBH to form more quickly, but the physical mechanism is still unclear. Black hole mergers and prolonged accretion are also thought to play a role in the growth of SMBH \citep[e.g.][]{gree20,mezc17}.  

Tidal disruption events (TDEs) occur when a star passes through the tidal radius of a massive black hole (MBH), where the tidal forces exceed the binding energy of the star causing it to be torn apart \citep{rees88}. Upon the encounter, about half of the stellar mass is expected to be captured and accreted by the black hole on a timescale of the order of $\sim$1 year \citep{hill75}, producing powerful outbursts with peak luminosities of up to 10$^{45}$ erg s$^{-1}$ \citep[e.g.][]{stru09}. Thus a black hole that was not undergoing active accretion, and therefore difficult to detect, can become very bright and easy to locate. 

A star may not, however, be totally disrupted. It is possible that only the outer regions of a (more massive) star are disrupted, leaving the stellar core to reform as a star \citep{rees88}. The fate of the star would then depend on the orbital parameters and the mass of the star, leaving it either unbound or in orbit around the MBH, either undergoing further (partial) disruption at periastron or orbiting at larger distances, where it may (repeatedly) intersect the accretion stream/disc, giving rise to outbursts \citep[e.g.][]{dai10}. Partial TDEs are expected to be observed more frequently than complete TDEs as the rate of encounters scales with the pericentre distance and because there is more chance to catch the TDE if the bursts repeat \citep{guil13}.

For a TDE of a solar type star, the tidal radius is inside the MBH event horizon if M$_{BH}\gtrsim$10$^8$ M$_\odot$ \citep{hill75}, so the TDE can not be observed. TDEs can then be used to find lower mass MBHs, where the number of MBHs increases towards lower masses, down to an unconstrained mass threshold \citep[e.g.][]{gree20}. Determining whether the trend extends to the IMBH range would help constrain the origin of SMBHs, where different distributions are expected depending on how the BH formed \citep[e.g.][]{gree20}. 

For TDEs where the black hole mass is $<$ a few $\times$ 10$^7$ M$_\odot$, the mass fallback rate is expected to significantly exceed the Eddington rate for weeks to years \citep[e.g.][]{stru09}. Observing these TDEs at their peak and modelling the spectra (and lightcurves) can help constrain the physical mechanism behind super-Eddington accretion and therefore understand its role in the growth of black holes. Observing the outburst duration gives insight into the mass accreted and the TDE rate gives a constraint on their role in SMBH growth.

However, only $\sim$100 TDEs have been observed to date\footnote{\url{https://tde.space/}} and amongst this small sample, the duration of the TDE outbursts is highly variable. Some TDEs detected in X-rays are observable for only half a year e.g. Swift J2058.4+0516 \citep{pash15}, whereas others, such as 3XMM J150052.0+015452 have stayed bright for more than a decade \citep{lin17b,lin22}. The origin of the range of outburst duration is not fully understood. It may depend on the viscosity of the matter \citep{rees88}, but a $>$10 year super-Eddington outburst would require an unphysically low viscosity. Alternatively, the outburst duration could be linked to the mass of the accreted star, or the circularisation of the debris stream may be inefficient due to weak general relativity effects, so that there is a high mass fallback rate \citep{chen18}. Increasing the sample size of X-ray TDEs and modelling the X-ray lightcurve will help determine the physical mechanism behind the duration of the outbursts and constrain the range of durations, essential for determining the mass accreted.

There are other open questions concerning TDEs. First, the emission was expected to be thermal due to radiative processes in the accretion flow \citep{rees88}, with a temperature peaking in the soft X-ray/UV, depending on the black hole mass. Whilst most TDEs do show very soft emission, peaking at energies $\lesssim$0.1 keV (see Sec.~\ref{sec:softTDEs}), a few have shown much harder X-ray emission, notably Swift J164449.3+573451 \citep{burr11} and Swift J2058.4+0516 \citep{cenk12}. The X-ray spectra are fitted with hard power laws and the emission extends to hundreds of keV.  Why some TDEs show harder spectra is unclear. It may be due to a jet pointing towards us \citep[e.g.][]{auch17}, but other mechanisms are also suggested \citep[e.g.][]{hryn16} and simply detecting the TDE after the peak as it moves into the low hard state can also explain the spectrum. If hard emission is due to a jet, determining the fraction of hard TDEs will give clues to the jet opening angle.  

It is not only in the X-ray that the emission properties are diverse. Some TDEs are not detected in X-rays at all and only at longer wavelengths (UV, optical, infra-red and/or radio), e.g. PTF-09ge \citep[][]{arca14}. The origin of the UV/optical TDE emission is unclear. It may be from reprocessing X-ray emission from the accretion disc by optically thick material surrounding it \citep[e.g.][]{guil13,roth16}, or from shocks between debris streams as they collide e.g. \citep{pira15}, or a combination of both \citep[e.g.][]{jian16,lu20}. Some TDEs are detected only in the X-ray and not in the optical, and some are seen in both \citep{auch17}, as anticipated by \cite{dai18}. More recently, a TDE  was also observed in the radio domain, but not in the optical nor in X-rays \citep{matt18}.  What is the origin of this diversity? It may simply be due to the viewing angle. Alternatively it could be caused by dust obscuration, or an as yet unknown physical mechanism. 

Recently another feature has been associated with some TDEs, quasi periodic eruptions \citep[QPEs,][]{mini19}. They are seen in the X-ray lightcurve as the source tends towards quiescence. The first was identified by \cite{mini19} who saw the X-ray count rate increase by up to two orders of magnitude over 1 h every 9 h in the Seyfert 2 galaxy GSN 069. These eruptions show a fast spectral transition where the soft thermal emission heats up. Since then, another four systems have shown QPEs, RX J1301.9+2747 \citep{gius20}, eRO-QPE1 and eRO-QPE2 \citep{arco21} and tentatively XMMSL1 J024916.6-041244 \cite{chak21}. However, \cite{arco22} showed that the time between bursts is not always quasi-periodic, with bursts sometimes arriving in pairs. Further, the burst start and peak times vary for different energies.

One theory is that QPEs are due to extreme mass-ratio inspirals (EMRIs), where a stellar mass object spirals towards the MBH due to energy loss via gravitational waves \citep[e.g.][]{metz22}. More than one star could be circling the MBH, accounting for pairs of bursts. Alternatively, an inspiralling object could repeatedly impact an accretion disc, possibly formed through a TDE, creating QPEs \citep{dai10}, but the mechanism is not yet satisfactorily explained. Discovering and observing new systems will help probe the QPE phenomenon. Here we discuss methods to find QPEs, discuss their nature and outline future work. 

\section{The XMM-Newton catalogues}
\label{sec:4XMM}

The XMM-Newton Survey Science Centre (XMM-SSC) \citep{wats01} has developed much of the {\it XMM-Newton} Science Analysis System (SAS) in collaboration with the ESA Science Operations Centre (SOC) \citep{gabr04}. The SAS enables the reduction and analysis of {\it XMM-Newton} data and is used in the pipeline to perform standardised routine processing of the {\it XMM-Newton} data. The XMM-SSC, in collaboration with the SOC, also produces catalogues of XMM-Newton detections. The detection catalogues made with data from the three EPIC cameras have been designated successively 1XMM, 2XMM, 3XMM and 4XMM \citep{webb20,rose16}, with incremental versions denoted -DR (Data Release). The latest version is 4XMM-DR12 (July 2022) and includes an extra year of data with respect to 4XMM-DR11. 4XMM-DR12 contains 939270 X-ray detections ($\geq$ 3 $\sigma$) which relate to 630347 unique X-ray sources from 12712 observations. 4XMM-DR12 covers 1283$^2$ degrees of sky, with $\geq$1 ks exposure. Some regions have been pointed 84 times. 9\% of detections are extended. Spectra and time series are extracted for detections with $\geq$100 EPIC counts, i.e. for 36\% of the catalogue. The median positional uncertainty is 1.57" ($\sigma\sim$1.43"). The median fluxes are $\sim$5.2$\times$10$^{-15}$ erg cm$^{-2}$ s$^{-1}$ (0.2-2.0 keV) and $\sim$1.2$\times$10$^{-14}$ erg cm$^{-2}$ s$^{-1}$ (2-12 keV).

The XMM-SSC also provides a catalogue created from overlapping observations. Stacking provides longer effective exposure times, resulting in better source parameters and higher sensitivity \citep{trau20}. 4XMM-DR12s, is made from 1620 groups from 9355 observations. Most stacks are composed of 2 observations and the largest has 372. 4XMM-DR12s contains 386043 sources, 298626 with several observations. The median source fluxes are $\sim$2.5$\times$10$^{-15}$ erg cm$^{-2}$ s$^{-1}$ (0.2-2.0 keV) and $\sim$6.8$\times$10$^{-15}$ erg cm$^{-2}$ s$^{-1}$ (2.0-12.0 keV). 6868 good quality sources have long-term variability.

The catalogues provide up to 336 columns of data that include identifiers/coordinates, observation date/time and observing mode, exposure and background information in the full 0.2-12.0 keV band (Band 8) and the five sub-bands, Band 1 (0.2-0.5 keV), Band 2 (0.5-1.0 keV), Band 3 (1.0-2.0 keV), Band 4 (2.0-4.5 keV) and Band 5 (4.5-12.0 keV). Information on the source extent, counts, fluxes and rates in all bands, hardness ratios (providing rudimentary spectral information), the reliability of a detection/source, quality flags and variability information. The slimline version has one row per source (rather than per detection) and a reduced number of columns.

The catalogues are provided as Flexible Image Transport System (FITS) or Comma Separated Values (CSV) files via the XMM-SSC webpages\footnote{\url{http://xmmssc.irap.omp.eu/}} and the XMM-Newton Science Archive (XSA)\footnote{\url{http://www.cosmos.esa.int/web/xmm-newton/xsa}}. The catalogue can also be queried through the XCATDB\footnote{\url{http://xcatdb.unistra.fr/4xmmdr12}}, HEASARC\footnote{\url{http://heasarc.gsfc.nasa.gov/W3Browse/xmm-newton/xmmssc.htm}} and the IRAP catalogue server\footnote{\url{http://xmm-catalog.irap.omp.eu/}}.

Complimentary catalogues are also provided, notably the XMM-Newton Optical Monitor (OM) Serendipitous Ultra-violet Source Survey Catalogue \citep[XMM-SUSS,][]{page12}. The latest version, XMM-SUSS5 contains 8.86 million detections (5.97 million unique sources) in different UV (UVW2, UVM2 and UVW1) and optical (U, B, and V) bands. Also provided is a catalogue of detections made with the pn camera during {\it XMM-Newton} slews between observations \citep{saxt08}. The slew survey catalogue (XMMSL1) covers a greater fraction of the sky (14\%) than the pointed observations, but with a shallower flux limit of 1.2 $\times$10$^{-12}$ erg cm$^{-2}$ s$^{-1}$ (0.2-12.0 keV). The first catalogue contains 2692 clean sources. To date 85\% of the sky has been observed \citep{saxt21}.

\section{Tidal disruption events discovered with XMM-Newton}

Many TDEs have been followed-up using {\it XMM-Newton} \citep[e.g.][]{auch17}, but thanks to observations spanning $>$22 years, reasonable sky coverage with repeated observations and good sensitivity in the X-ray domain, it offers a unique opportunity to discover TDEs. Here we review a range of TDEs discovered using {\it XMM-Newton} data. For a more complete review of X-ray TDEs, see \cite{saxt21}.

\begin{figure*}[t]
	\centerline{\includegraphics[width=185mm,height=18pc]{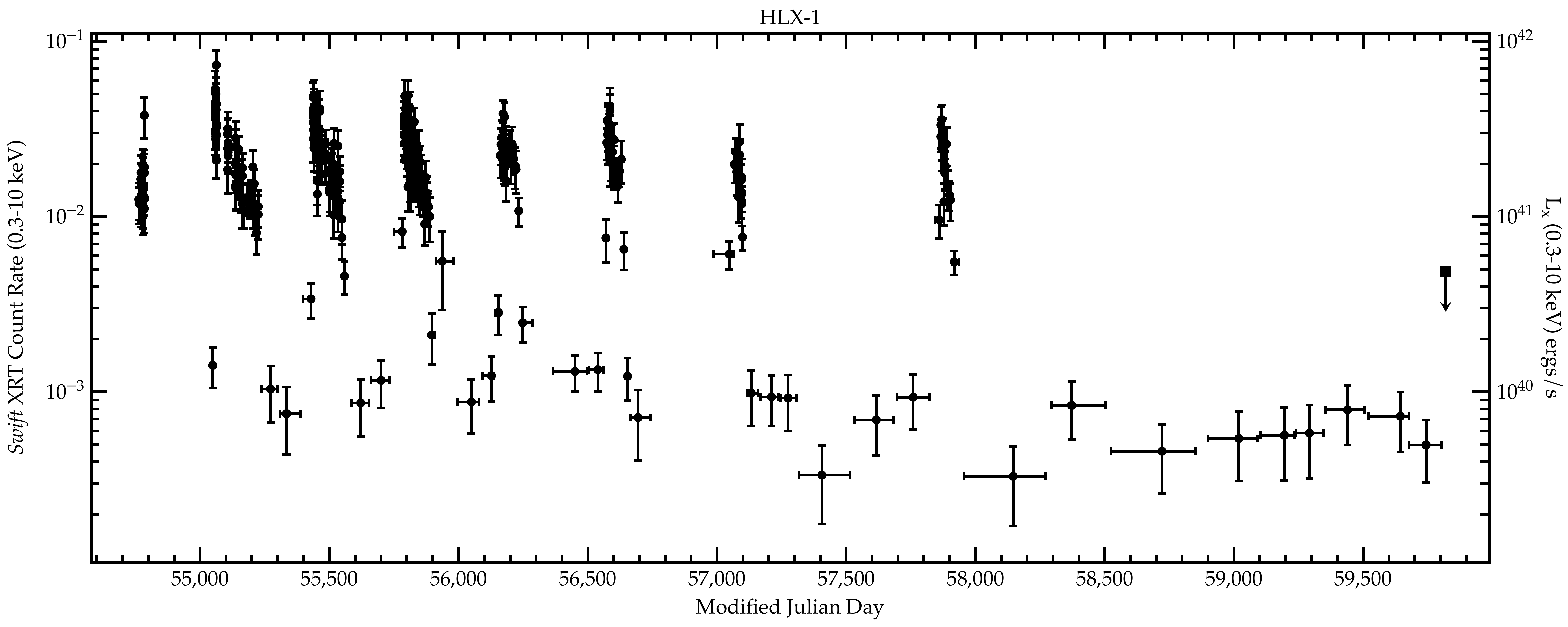}}
	\caption{The recent Swift X-ray lightcurve of HLX-1, spanning fourteen years, from 2008-2022\label{fig:HXL1lc}}
\end{figure*}

\subsection{Soft tidal disruption events}
\label{sec:softTDEs}

Thanks to two {\it XMM-Newton} EPIC observations showing a flux increase of a factor $\sim$5 in 7 months and showing no {\it Rosat} counterpart at an upper limit of almost a factor 10 fainter, a decade previously, 2XMMi J184725.1-631724 was shown to be a soft TDE, with a blackbody temperature increasing from 57.8 to 78.2 eV \citep{lin11}. 4 years later the flux had decreased by a factor of $>$12. The host galaxy IC 4765-f01-1504, situated at z=0.0353, was shown to be inactive using GMOS/Gemini South spectroscopy. Modelling the blackbody emission with a kerrbb model in {\it XSpec} indicated a fairly low black hole mass of 0.06 - 4 $\times$ 10$^6$ M$_\odot$ \citep{lin11}. 

Similarly, 3XMM J152130.7+074916 has only one {\it XMM-Newton} detection but another upper limit shows flux variability of almost a factor 1000 \citep{lin15}. It is consistent with the centre of the galaxy SDSS J152130.72+074916.5 at z = 0.17901 (866
Mpc). The blackbody temperature is 170 eV and a rest-frame 0.24-11.8 keV unabsorbed luminosity of $\sim$5 $\times$ 10$^{43}$ erg s$^{-1}$ \citep{lin15}, similar to 2XMMi J184725.1-631724. The mass estimate is similar to that for 2XMMi J184725.1-631724, indicating that they must have reached similar Eddington ratios if observed close to the peak.

Another object is 3XMM J141711.1+522541 in the inactive S0 galaxy SDSS J141711.07+522540.8 at z = 0.418 \citep{lin16}, with a similar luminosity at peak and a blackbody temperature ranging from $\sim$113-130 eV. Supposing that the source reached the Eddington luminosity around the peak, the black hole mass would be $\sim$10$^5$ M$_\odot$ \citep{lin16}.  It is believed to reside in the centre of an ultra compact dwarf galaxy.

These systems are similar to SDSS J120136.02+300305.5 \citep{saxt12}, but clearly different to XMMSL1 J074008.2-853927 \cite{saxt17} which has a similar blackbody, but is dominated in the X-ray by a power law tail with $\Gamma\sim2$. This TDE appears to be between the soft TDEs described above and the relativistic, hard TDEs described in Sec.~\ref{sec:intro}. More observations of these intermediate TDEs may help us understand why some spectra are soft and others hard.

\subsection{Intermediate mass black holes in TDEs}

Whilst the MBHs in the TDEs discussed in Sec.~\ref{sec:softTDEs} have low masses, 3XMM J215022.4$-$055108 \citep{lin18,lin20} appears to be similar, reaching a peak luminosity of $\sim$1 $\times$ 10$^{43}$ erg s$^{-1}$ (0.2-10.0 keV) with a disc temperature of $\sim$140-280 eV,  but it has spectra that reveal a high black hole spin rate of 0.92-1 and a possible mass range of 5.3-12 $\times$ 10$^4$ M$_\odot$, making it an excellent candidate for an intermediate mass black hole. The low mass estimate is confirmed by \cite{chen18} who estimate 7.1 $\times$ 10$^4$ M$_\odot$ through modelling the lightcurve.

\subsection{Long TDEs}

\cite{auch17} used the T$_{90}$ value, traditionally used for $\gamma$-ray bursts, to estimate the duration of TDEs. The T$_{90}$ is the time period over which 90\% of the total fluence is recorded. T$_{90}$ can quantify the TDE duration, however, it depends on the sensitivity and energy range and response of the instruments used, as well as the duration a source is followed. Further, with poorly sampled lightcurves, it can be difficult to determine the duration. Here we take a different approach for the X-ray TDEs, considering the time for the estimated peak luminosity to fall by a factor 100 assuming that the source fades following the approximation t$^{-5/3}$, to give an indication of the outburst duration. For the first three TDEs presented in Sec.~\ref{sec:softTDEs} we determine 1.58 $\times$ 10$^8$, 1.30 $\times$ 10$^8$ and 2.6 $\times$ 10$^8$ s respectively. For comparison, \cite{auch17} determined a T$_{90}$ of 0.45 $\times$ 10$^8$ s for the second TDE, 3XMM J152130.7+074916.  In contrast, the TDE 3XMM J150052.0+015452 has a similar peak luminosity to the TDEs cited in Sec.~\ref{sec:softTDEs} and is found at a similar distance of z$\sim$0.145, with an equally soft black body of $\sim$300 eV \cite{lin17b}, but after 15 years it has only dropped in luminosity by a factor 3, making the outburst duration $\gg$ 4.73 $\times$ 10$^8$ s. It is predicted to decay by a factor 10 over the next 14 years \citep{lin22}, which suggests an outburst duration $>$ 9.15 $\times$ 10$^8$ s.  Modelling the X-ray spectra indicates that the source was in a super-Eddington state for about five years, before dropping to approximately Eddington luminosity, where it remains. The modelling implies that the MBH has a high spin ($>0.8$) and a mass of a few $\times$ 10$^5$ M$_\odot$ \citep{lin22} and that the disrupted star had a mass of $\sim$0.75 M$_\odot$.  Around 0.28 M$_\odot$ has been accreted, based on the lightcurve and the X-ray spectra \citep{lin22}. .

\subsection{TDEs showing long-term variability}

\subsubsection{Partial TDEs}
\label{sec:partialTDEs}

2XMM J011028.1-460421, commonly known as Hyper Luminous X-ray source 1 \citep[HLX-1, ][]{farr09} reaches a maximum luminosity of $\sim$1.3 $\times$ 10$^{42}$ erg s$^{-1}$ (0.2-10.0 keV) and has a mass of $\sim$10$^4$ M$_\odot$ \citep{gode12}, also making it an IMBH. It is highly variable, see Fig.~\ref{fig:HXL1lc}, showing eight outbursts since 2008 each with a factor $\sim$50 rise in luminosity, initially with a recurrence time of $\sim$1 year but increasing to $>$5 years. The bursts show soft blackbody spectra (kT$\sim$200 eV). In the fainter state, the spectra are best fitted with a hard power law ($\Gamma\sim$2.2) \citep{gode12}, similar to stellar mass black hole X-ray binaries, but with luminosities $\sim$1000 times brighter. Periodic accretion from a companion star in a highly elliptical orbit, which is tidally stripped as it approaches periastron is thought to explain the variability \citep{gode14,webb14}, which is consistent with that expected for a repeating, partial TDE \citep{macl13}. HLX-1's spectrum in the low state is clearly different from that observed for QPEs. We propose that HLX-1's outbursts may be produced in the same way as the QPEs, but the long orbit may provide time for the accretion disc to empty between periastron passages of the star, allowing it to enter the low/hard state.

HLX-1 is 8'' from the centre of the galaxy ESO 243-49, in the galaxy cluster Abell 2877. ESO 243-49, one of the more massive cluster galaxies, located close to the cluster centre means that it is likely to have suffered dynamical effects, such as interactions or accretion of other bodies, making it probable that HLX-1 stems from a minor merger with its host \citep{webb10,mape12,webb17}, although no evidence for a recent merger has been found \citep{musa15,webb17}.  Such a merger could be responsible for placing the companion star to the IMBH in its elliptical orbit. As the time between outbursts is becoming progressively longer, it appears that the star is becoming unbound, due to the tidal forces between the two components \citep[][]{gode14}.

\subsection{Quasi-periodic oscillations and eruptions}

 2XMM J123103.2+110648 \citep{lin13}, coincident with the centre of the galaxy SDSS J123103.24+110648.6, showed a peak X-ray luminosity of  $\sim$4 $\times$ 10$^{42}$ erg s$^{-1}$ (0.2-10.0 keV), with a multi-colour disc blackbody varying from $\sim$ 90-200 eV,  along with a fairly sinusoidal modulation of around 3.8 h (a quasi-periodic oscillation, QPO) at peak \citep{lin13}. Follow-up observations confirmed it was indeed a TDE \citep{lin17a}. 2XMM J123103.2+110648 strongly resembles the QPE source GSN 069 discovered in the {\it XMM-Newton} slew survey \citep{saxt08}, also believed to be a TDE \citep[e.g.][]{mini22}, showing no hard X-ray emission, strong, fast variability, no broad H$_\alpha$ or H$_\beta$ lines and a similar disc emission from a low mass MBH \citep{lin13,lin17a}. The nature of the very soft QPO is unclear. It may be due to a special accretion mode in TDEs \cite{lin13}. Indeed, \cite{mini19} stated that if QPEs are due to a disc-instability, limit-cycle oscillations should become low-amplitude QPOs as the mass transfer rate falls, due to the shrinking size of the unstable region. However, recent work suggests that QPEs are more likely to be associated with EMRIs, see Sec.~\ref{sec:intro}. We propose that there may be a relation between the very soft QPOs and QPEs, where the QPOs could arise from orbiting material (or the remains of a partially disrupted star) from a (partial) TDE, in a similar way to the QPEs that may be due to a partial TDE, leaving behind the stripped stellar core, repeatedly impacting the accretion disc/stream around the MBH.

\section{Future work}

As can be seen, the TDE phenomenon is highly varied, and is useful for finding the low mass MBH that could be the seeds of the SMBH, as well as for understanding the importance of mass accretion in the growth of SMBH. Finding more TDEs will help in understanding the demographics of IMBH. New TDEs, notably partial TDEs \citep[e.g.][]{liu22} will be discovered with new surveying facilities, such as {\it eROSITA} or the {\it Vera Rubin observatory} in the optical, or the {\it Square Kilometre Array} (SKA) in the radio, which will be able to detect almost any quiescent IMBH in our Galaxy.  Finally, moving to gravitational waves, future facilities, such as {\it LISA} will also be able to detect MBH mergers and EMRI \citep{bara15}. 

However, using optimised searches to systematically exploit all existing X-ray data to increase the sky area searched and the baseline of X-ray observations, along with contemporary optical and UV data from the OM, for example, should also reveal many new transients, including TDEs (Quintin et al. to be sub.). Alternatively, machine learning techniques can be used to identify TDEs in large X-ray catalogues \citep[e.g.][]{tran22} or dedicated automated searches for bursting sources (Gupta et al. to be sub.) could reveal new QPEs, so that it will finally be possible to understand their nature and possible relation with TDEs and QPOs.

%\backmatter

\section*{Acknowledgments}
This work was funded in part by the CNES \& the Horizon 2020 research and innovation programme under grant agreement n°101004168, the XMM2ATHENA project.

%\subsection*{Author contributions}

%\subsection*{Financial disclosure}

%\subsection*{Conflict of interest}

%The authors declare no potential conflict of interests.

%\nocite{*}% Show all bib entries - both cited and uncited; comment this line to view only cited bib entries;
\bibliography{Webb}

%\section*{Author Biography}
%(if applicable)

%\begin{biography}{\includegraphics[width=60pt,height=70pt,draft]{empty}}{\textbf{Author Name.} This is sample author biography text this is sample author biography text this is sample author biography text this is sample author biography text this is sample author biography text this is sample author biography text this is sample author biography text this is sample author biography text this is sample author biography text .}
%\end{biography}

\end{document}